\documentclass[10pt, conference, letterpaper]{IEEEtran}

\usepackage{subcaption} 
\usepackage{stmaryrd}
\usepackage{verbatim}
\usepackage{listings}
\usepackage{array}
\usepackage{multirow}
\usepackage{tikz}
\usepackage{bbm}
\usetikzlibrary{fit,arrows,calc,spy,decorations.pathreplacing,shapes, patterns}
\usepackage{graphicx} 
\usepackage{stmaryrd} 
\usepackage{color}
\usepackage{amsmath} 
\usepackage[utf8]{inputenc}
\usepackage{amssymb} 
\usepackage{epstopdf}

\title{Degree-based Outlier Detection within IP Traffic Modelled as a Link Stream}

\author{
\IEEEauthorblockN{Audrey Wilmet\IEEEauthorrefmark{1}, Tiphaine Viard\IEEEauthorrefmark{3}, Matthieu Latapy\IEEEauthorrefmark{1}, Robin Lamarche-Perrin\IEEEauthorrefmark{2}}
\IEEEauthorblockA{\IEEEauthorrefmark{1}Sorbonne Universit{\'e}, CNRS, Laboratoire d'Informatique de Paris 6, LIP6, F-75005 Paris, France}
\IEEEauthorblockA{\IEEEauthorrefmark{2}Institut des Systèmes Complexes de Paris Île-de-France, ISC-PIF, UPS 3611, Paris, France\\
Email: firstname.lastname@lip6.fr }
\IEEEauthorblockA{\IEEEauthorrefmark{3}Discrete Optimization Unit, Riken AIP, Tokyo, Japan, tiphaine.viard@riken.jp}
}

\begin{document}
\maketitle

\begin{abstract}
This paper aims at precisely detecting and identifying anomalous events in IP traffic. To this end, we adopt the link stream formalism which properly captures temporal and structural features of the data. Within this framework, we focus on finding anomalous behaviours with respect to the degree of IP addresses over time. Due to diversity in IP profiles, this feature is typically distributed heterogeneously, preventing us to directly find anomalies. To deal with this challenge, we design a method to detect outliers as well as precisely identify their cause in a sequence of similar heterogeneous distributions. We apply it to several MAWI captures of IP traffic and we show that it succeeds in detecting relevant patterns in terms of anomalous network activity.
\end{abstract}

\section{Introduction}
Temporal and structural features of IP traffic are and have been for several years the subject of multiple studies in various fields. A significant part of this research is devoted to detecting statistically anomalous traffic subsets referred to as anomalies, events or outliers. Their detection is particularly important since, in addition to a better understanding of IP traffic characteristics, it could prevent attacks against on-line services, networks and information systems. \\
Due to the temporal and structural nature of IP traffic, developed methods can be classified into two categories: those based on signal processing \cite{barford2002signal, borgnat2009seven}, and those based on graph theory \cite{latapy2013detecting,Xu2014BehaviorAO}. However, methods within these areas lead to a loss of information: by considering interactions as a signal, the structure is aggregated; with graphs, the temporal order of interactions is lost. Hence, existing methods struggle to identify subtle outliers, which are abnormal both in time and structure. Moreover, when they succeed, the loss of information leads to a decrease of accuracy. In this paper, we model IP traffic as a link stream which fully captures both the temporal and the structural nature of traffic \cite{latapy2017stream,viard2017discovering}. Then, we introduce a method in order to detect subtle outliers and precisely identify which IP addresses and instants caused it. \\
More specifically, a link stream $L$ is defined as a set of instants $T$, a set of nodes $V$ (IP addresses) and a set of interactions $E$ (communication between IP addresses over time). Within this framework, we focus on one key property: the  degree of nodes. This feature is highly heterogeneous, which raises challenges for its use in outlier detection, but it is stable over time. Our method takes advantage of this temporal homogeneity: it divides the link stream into time slices and then performs outlier detection to find time slices which exhibit unusual number of nodes having a degree within specific degree classes. Then, in order to isolate responsible IP addresses and instants on which they behave unexpectedly, we design an identification method based on an iterative removal of previously detected events. Finally, we validate our method by showing that these event removals do not significantly alter the underlying normal traffic. \\
This paper is an extended version of the work published in \cite{wilmet2018degree}. In this contribution, special attention is paid to the importance of parameters involved in the method such as time slice and degree class sizes. Moreover, in addition to our previous work in which we evaluated our method on a one-hour long IP traffic trace of June $2013$, we apply it on two other datasets: a one-day long IP traffic trace of June $2013$ and a fifteen-minutes long IP traffic trace of November $2018$ which possesses a list of abnormal events indexed by MAWILab to which we compare our results \cite{fontugne2010mawilab}. \\ The paper is organised as follows. We overview the related work in Section \ref{sec:related_work} and present our contributions in Section \ref{sec:contributions}. We introduce IP traffic modelling as a link stream and the degree definition in Section \ref{sec:notions}. In Section \ref{sec:degree}, we describe our goals and the challenges they raise. This leads to the development of our method to detect events in Section \ref{sec:detection} and to identify them in Section \ref{sec:identification}. We discuss our results in Section \ref{sec:Validation}. Subsequently, we apply our method on other datasets in Section \ref{sec:datasets}. Then, we investigate the influence of different time slice sizes and different class constructions in Section \ref{sec:para}. We conclude in Section \ref{sec:conclusion}.

\section{Related Work}
\label{sec:related_work}

Techniques for anomaly detection in IP traffic are extremely diverse. Among those, methods using principal component analysis \cite{lakhina2004diagnosing, ringberg2007sensitivity}, machine learning \cite{williams2006preliminary}, data mining \cite{lee1998data}, signal analysis \cite{barford2002signal, borgnat2009seven} and graph-based techniques have been proposed. In this paper, we focus our related work on methods based on dynamic graphs. In this domain, authors traditionally study a sequence of graphs $\{G_i\}_{i=1..k}$, such that each snapshot $G_i=G_{i\tau .. (i+1)\tau}$ contains interactions aggregated over time window $T_i=[i\tau,(i+1)\tau[$. Then, they attribute an abnormality score to each snapshot $G_{i}$ by comparing it to others. This problem has been approached in various ways depending on the definition of the abnormality score (for surveys, see \cite{akoglu2015graph,ranshous2015anomaly}).\\

Compression-based abnormality scores analyse the evolution of the encoding cost of each graph to detect anomalies. Sun \textit{et al.} \cite{sun2007graphscope} and Duan \textit{et al.} \cite{duan2009community} group similar consecutive snapshots into a chain. If the adding of a graph greatly increases the description length of the chain, the corresponding snapshot is considered as abnormal. Chakrabarti \textit{et al.} \cite{chakrabarti2004autopart} use a similar technique but apply it on clusters of nodes to find abnormal links. \\
Other approaches use tensor decomposition. Ide \textit{et al.} \cite{ide2004eigenspace} and Ishibashi \textit{et al.} \cite{ishibashi2010detecting} build a past activity vector from the main eigenvectors associated to all snapshots within a given time window. Then, snapshot $G_i$ is identified as abnormal when the distance between the past activity vector and $G_i$'s main eigenvector exceeds a certain threshold. Akoglu \textit{et al.} \cite{akoglu2010event} use a similar technique in which the past activity vector includes nodes local features.\\
Outliers can also be found by studying communities evolution. Aggarwal \textit{et al.} \cite{aggarwal2016outlier} find anomalous snapshots by comparing their clustering quality. Gupta \textit{et al.} \cite{gupta2012integrating,gupta2012community} calculate the variation in the probability of belonging to a community for each node between two consecutive snapshots. Nodes for which the variation deviates significantly from the average variation of nodes within the same community are considered abnormal. Chen \textit{et al.} \cite{chen2012community} and Araujo \textit{et al.} \cite{araujo2014com2}, in turn, detect abnormal communities among clusters which unusually increase, merge, decrease or split.\\
Finally, a significant amount of work quantify the distance between snapshots using graph features. Pincombe \textit{et al.} \cite{pincombe2005anomaly} and Papadimitriou \textit{et al.} \cite{papadimitriou2010web} define a series of topological aggregated features to compare snapshots. Berlingerio \textit{et al.} \cite{berlingerio2012netsimile} use the moments of an egonet feature -- \textit{e.g.}, degree, clustering coefficient -- calculated on each node. Saxena \textit{et al.} \cite{saxena2016leveraging} use a similar method but decompose each snapshot into $k$ cores to consider global features as well. Schieber~\textit{et~al.} \cite{schieber2017quantification} use the Jensen-Shannon divergence and a measure of the heterogeneity of each graphs in terms of connectivity distance between nodes. Finally, Mongiovi \textit{et al.} \cite{mongiovi2013netspot} find clusters of anomalous links by calculating, for each link, its probability to have a given weight according to its usual behaviour.\\

However, these techniques lead to a loss of information: by reducing interactions into a sequence of graphs, the links order of arrival within a time window is lost. To overcome this issue, other work propose to improve these methods by introducing sequences of augmented graphs. For instance, Casteigts \textit{et al.} \cite{casteigts2012time} and Batagelj \textit{et al.} \cite{batagelj2016algebraic} use graphs in which links are labelled with their instants of occurrence. Likewise, in \cite{asai2014network,wehmuth2015unifying} and \cite{whitbeck2012temporal}, authors use causal graphs in which two nodes are linked together if there is a causal relationship between them. In this paper, we adopt a new perspective. We consider temporal interactions as a separate object called a link stream, using the formalism developed by Latapy \textit{et al.} \cite{latapy2017stream}. While methods for covering relevant and subtle events often go hand in hand with very complicated features, we show that we can find relevant structural and temporal outliers, as well as gain accuracy, with a very simple feature defined in the link stream formalism: the instantaneous degree of nodes over time. \\

Other authors detect outliers using this modelling. Yu \textit{et al.} \cite{yu2013anomalous} calculate the main eigenvector of the ego-network of each node and find abnormal nodes among those experiencing a sudden change in the amplitude and/or direction of their vectors. Manzoor \textit{et al.} \cite{manzoor2016fast} use a similar technique. They store the link stream in a sketch built from the ego-networks patterns of each node and label a new edge as abnormal if the difference between the sketch before its arrival and the one after is significant. Ranshous \textit{et al.} \cite{ranshous2016scalable} also use sketches. They store the link stream in a Count Min sketch that approximates the frequency of links and nodes. From this sketch, they assign an abnormality score to each link $(u, v, t)$ based on prior occurrences, preferential attachment and mutual neighbours of nodes $u$ and $v$. Eswaran \textit{et al.} \cite{eswaran2018sedanspot} also rely on approximations and attribute a score to every new edge arriving in the stream by relying on a sub-stream $L'$ sampled from past edges. If the new edge connects parts of $L'$ which are sparsely connected, then it is considered as abnormal. Finally, Viard \textit{et al.} \cite{viard2017discovering} find anomalous bipartite cliques using the link stream formalism developed in \cite{latapy2017stream}.\\

A large proportion of the methods cited above is devoted to find globally anomalous instants (as abnormal snapshots). Among those extracting local features on nodes or links, either authors use similarity functions which aggregate local information, or they rely on approximations as samples or sketches. In the first case, instants are abnormal based on their local patterns but information about which sub-graphs are responsible is lost. In the second case, approximations allow a fast processing but lead to a decrease of accuracy. In contrast, our method identify abnormal couples $(t,v)$ exactly, without any information loss and still exhibits fast and efficient processing.

\section{Contributions}
\label{sec:contributions}

We model IP traffic with a link stream and study one of its most important properties, the degree of nodes over time. We show that, although this property follows a very heterogeneous distribution that is hard to model, this distribution is stable over time. We then design a method that exploits the stability of this heterogeneity for anomaly detection, and may be applied in various such situations. This method first splits traffic into time slices and computes the degree distribution in each slice. By comparing these distributions, the method then points out degree classes and time slices such that having a degree in this class during this slice is anomalous. Using this information, we identify IP addresses and time periods involved in anomalies, as well as the corresponding traffic. By removing this traffic from the original data, we validate our identification by noticing that we turn back to a normal traffic with respect to the degree. We illustrate the method and its outcome on MAWI public IP traffic.

\section{Traffic Modelled as a Link Stream}
\label{sec:notions}

IP traffic consists of packet exchanges between IP addresses. We use here one hour of IP traffic capture from the MAWI archive\footnote{http://mawi.wide.ad.jp/mawi/ditl/ditl2013/ \cite{kato1999internet}} on June $25^{\text{th}}$, 2013, from 00:00 to 01:00. We denote this trace by a set $\mathcal{D}$ of triplets such that $(t,u,v)\in \mathcal{D}$ indicates that IP addresses $u$ and $v$ exchanged a packet at time $t$. The set $\mathcal{D}$ contains $83,386,538$ triplets involving $1,157,540$ different IP addresses. \\

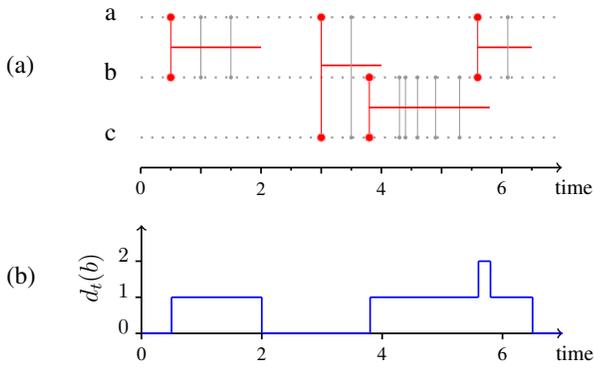
\begin{figure}

\vspace*{0.1cm}
\scalebox{0.8}{
\begin{tikzpicture}
\node[label={\large{(a)}}] (1) at (-2,4.75){};

\tikzstyle{nod2} = [circle, draw,red!40, scale=.8, fill=red, scale=0.5];
\tikzstyle{nod1} = [circle, draw, scale=.3,fill=black!5!white, scale=0.5];

\node[label={\large{a}}] (1) at (-.5,5.7){};
\node[label={\large{b}}] (1) at (-.5,4.7){};
\node[label={\large{c}}] (1) at (-.5,3.7){};

\draw[loosely dotted, black!40, line width=1pt] (0,6) -- (7,6);

\draw[loosely dotted, black!40, line width=1pt] (0,5) -- (7,5);

\draw[loosely dotted, black!40, line width=1pt] (0,4) -- (7,4);

\draw[thick,->] (0,3.5) --  (7,3.5);
\node[label={\small{0}}] (1) at (0,2.8){};
\node[label={\small{2}}] (1) at (2,2.8){};
\node[label={\small{4}}] (1) at (4,2.8){};
\node[label={\small{6}}] (1) at (6,2.8){};
\node[label={time}] (1) at (7.2,2.8){};
\draw[thick] (0,3.5) --  (0,3.4);
\draw[thick] (0.5,3.5) --  (0.5,3.45);
\draw[thick] (1,3.5) --  (1,3.4);
\draw[thick] (1.5,3.5) --  (1.5,3.45);
\draw[thick] (2,3.5) --  (2,3.4);
\draw[thick] (2.5,3.5) --  (2.5,3.45);
\draw[thick] (3,3.5) --  (3,3.4);
\draw[thick] (3.5,3.5) --  (3.5,3.45);
\draw[thick] (4,3.5) --  (4,3.4);
\draw[thick] (4.5,3.5) --  (4.5,3.45);
\draw[thick] (5,3.5) --  (5,3.4);
\draw[thick] (5.5,3.5) --  (5.5,3.45);
\draw[thick] (6,3.5) --  (6,3.4);
\draw[thick] (6.5,3.5) --  (6.5,3.45);

\node[nod1,black!40!white] (0) at (1,6) {};
\node[nod1,black!40!white] (1) at (1,5) {};
\path[-,black!40!white] (0) edge (1);
\node[nod1,black!40!white] (0) at (1.5,6) {};
\node[nod1,black!40!white] (1) at (1.5,5) {};
\path[-,black!40!white] (0) edge (1);
\node[nod2] (0) at (0.5,6) {};
\node[nod2] (1) at (0.5,5) {};
\path[-,red] (0) edge (1);
\draw[thick,red](0.5,5.5) --(2,5.5);

\node[nod1,black!40!white] (0) at (4.3,5) {};
\node[nod1,black!40!white] (1) at (4.3,4) {};
\path[-,black!40!white] (0) edge (1);
\node[nod1,black!40!white] (0) at (4.4,5) {};
\node[nod1,black!40!white] (1) at (4.4,4) {};
\path[-,black!40!white] (0) edge (1);
\node[nod1,black!40!white] (0) at (4.6,5) {};
\node[nod1,black!40!white] (1) at (4.6,4) {};
\path[-,black!40!white] (0) edge (1);
\node[nod1,black!40!white] (0) at (4.9,5) {};
\node[nod1,black!40!white] (1) at (4.9,4) {};
\path[-,black!40!white] (0) edge (1);
\node[nod1,black!40!white] (0) at (5.3,5) {};
\node[nod1,black!40!white] (1) at (5.3,4) {};
\path[-,black!40!white] (0) edge (1);
\draw[thick,red](3.8,4.5) --(5.8,4.5);
\node[nod2] (0) at (3.8,5) {};
\node[nod2] (1) at (3.8,4) {};
\path[-,red] (0) edge (1);

\node[nod1,black!40!white] (0) at (6.1,6) {};
\node[nod1,black!40!white] (1) at (6.1,5) {};
\draw[thick,red](5.6,5.5) --(6.5,5.5);
\path[-,black!40!white] (0) edge (1);
\node[nod2] (0) at (5.6,6) {};
\node[nod2] (1) at (5.6,5) {};
\path[-,red] (0) edge (1);

\node[nod1,black!40!white] (0) at (3.5,6) {};
\node[nod1,black!40!white] (1) at (3.5,4) {};
\draw[thick,red](3,5.2) --(4,5.2);
\path[-,black!40!white] (0) edge (1);
\node[nod2] (0) at (3,6) {};
\node[nod2] (1) at (3,4) {};
\path[-,red] (0) edge (1);

\end{tikzpicture}
}

\vspace*{0.1cm}
\scalebox{0.8}{
\begin{tikzpicture}

\node[label={\large{(b)}}] (1) at (-2,3.7){};

\tikzstyle{nod2} = [circle, draw,red!40, scale=.8, fill=red!40, scale=0.5];
\tikzstyle{nod1} = [circle, draw, scale=.3,fill=black!5!white, scale=0.5];
\node[label={[rotate=90]\large{$d_t(b)$}}] (1) at (-0.5,4){};

\node[label={$1$}] (1) at (-.3,3.55){};
\node[label={$2$}] (1) at (-.3,4.15){};
\node[label={$0$}] (1) at (-.3,2.95){};

\draw[thick,->] (0,3.2) --  (0,5);
\draw[thick,->] (-0.1,3.2) --  (7,3.2);

\draw[thick,-,blue] (0,3.2) --  (0.5,3.2);
\draw[thick,-,blue] (0.5,3.2) --  (0.5,3.8);
\draw[thick,-,blue] (0.5,3.8) -- (2,3.8);
\draw[thick,-,blue] (2,3.2) -- (2,3.8);

\draw[thick,-,blue] (2,3.2) --  (3.8,3.2);
\draw[thick,-,blue] (3.8,3.2) --  (3.8,3.8);
\draw[thick,-,blue] (3.8,3.8) -- (5.6,3.8);
\draw[thick,-,blue] (5.6,3.8) -- (5.6,4.4);
\draw[thick,-,blue] (5.6,4.4) -- (5.8,4.4);
\draw[thick,-,blue] (5.8,3.8) -- (5.8,4.4);
\draw[thick,-,blue] (5.8,3.8) -- (6.5,3.8);
\draw[thick,-,blue] (6.5,3.2) -- (6.5,3.8);

\draw[thick,-,blue] (6.5,3.2) --  (7,3.2);

\node[label={\small{0}}] (1) at (0,2.5){};
\node[label={\small{2}}] (1) at (2,2.5){};
\node[label={\small{4}}] (1) at (4,2.5){};
\node[label={\small{6}}] (1) at (6,2.5){};
\node[label={time}] (1) at (7.2,2.5){};
\draw[thick] (0,3.2) --  (0,3.1);
\draw[thick] (2,3.2) --  (2,3.1);
\draw[thick] (4,3.2) --  (4,3.1);
\draw[thick] (6,3.2) --  (6,3.1);

\draw[thick] (-0.1,3.8) --  (0,3.8);
\draw[thick] (-0.1,4.4) --  (0,4.4);
\end{tikzpicture}
}
\caption{\small{\textbf{Link stream for the modelling of IP traffic -} (a) Example of a link stream $ L=(T, V, E)$ formed from the set of triplets $\mathcal{D}=\{(1,a,b),(1.5,a,b),(3.5,a,c),(4.3,b,c),$ $(4.4,b,c),(4.6,b,c),(4.9,b,c),(5.3,b,c),(6.1,a,b)\}$: $T=[0,7[$, $V=\{a,b,c\}$, $E=\left( \,\left[ 0.5,2\right[ \:\cup\: \left[ 5.5,6.5\right[\, \right)\times \left\lbrace ab \right\rbrace$ $\:\cup\: \left[ 3,4\right[ \times \left\lbrace ac \right\rbrace \:\cup\: \left[ 3.8,5.8\right[ \times \left\lbrace bc \right\rbrace$. In the example, $a$ interacts with $b$ from $t_1=0.5$ to $t_2=2$. (b) Time evolution of the degree of node $b$. }}
\label{fig:stream_graph}
\end{figure}

\noindent We model this traffic as a link stream $L$ in order to capture its structure and dynamics \cite{latapy2017stream}. Nodes are IP addresses involved in $\mathcal{D}$ and two nodes are linked together from time $t_1$ to time $t_2$ if they exchanged at least one packet every second within this time interval. Formally, $L=(T,V,E)$ is defined by a time interval $T \subset \mathbb{R}$, a set of nodes $V$ and a set of links $E \subseteq T \times V \otimes V $ where $V \otimes V $ denotes the set of unordered pairs of distinct elements of $V$, denoted by $uv$ for any $u$ and $v$ in $V$ (thus, $uv\in V\otimes V$ implies that $u,v\in V$ and $u\neq v$, and we make no distinction between $uv$ and $vu$). If $(t, uv) \in E$, then $ u $ and $ v $ are linked together at time $t$. In our case, we take $E=\cup_{(t,u,v)\in \mathcal{D}}\: [t-\frac{\Delta}{2}, t+\frac{\Delta}{2}] \times \{uv\}$ with $\Delta=1\text{s}$. Other choices can be made. For instance, we can set a value of $\Delta$ that is different for each $uv$, or each link $(t, uv)$, using external knowledge. We can also use a value of $\Delta$ that changes over time (see for instance the work of Léo \textit{et al.} \cite{leo2019non}). These operations are depicted in Figure \ref{fig:stream_graph}.a.\\

\noindent The degree of $(t,v)\in T\times V$, denoted by $d_t(v)$, is the number of distinct nodes with which $v$ interacts at time $t$: $$d_t(v)=|\lbrace u\,:\,(t,uv)\in E\rbrace|.$$ Figure \ref{fig:stream_graph}.b shows the degree of node $b$ over time. Notice that the degree of $b$ is not its number of exchanged packets over time; it accounts for its number of distinct neighbours over time.

\section{Heterogeneity of Degrees}
\label{sec:degree}
In order to find outliers in a link stream using the degree, we first need to characterize the normal behaviour of the set of observations $\mathcal{O}=\{d_t(v) \: :\: (t,v) \in T\times V\}$. Then, an outlier is a couple $(t,v) \in T\times V$ which has a significantly different degree from others.\\

\begin{figure}[ht]
\hspace*{0.3cm}
\scalebox{0.6}{
\input{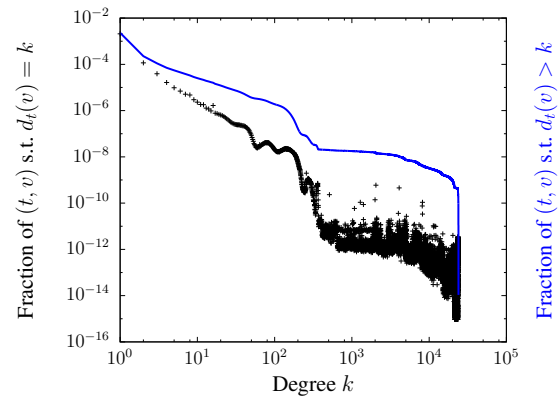}
}
\caption{\small{\textbf{Degree distribution and complementary cumulative degree distribution in  $L$.} For all $(t,v) \in T\times V$, we compute the degree $d_t(v)$ and plot the distribution of the set of values $\mathcal{O}=\{d_t(v) \: :\: (t,v) \in T\times V\}$. The fraction expresses the probability to draw a time instant $t\in T$ and a node $v\in V$ such that $d_t(v)=k$.}}
\label{fig:global_degree}
\end{figure}

\noindent For this purpose, we call degree distribution of $L$ the fraction $f(k)$ of couples $(t,v) \in T\times V$ for which $d_t(v)=k$, for all $k\in \mathbb{N}$: 
$$f(k)=\frac{|\{(t,v)\in T \times V \::\: d_t(v)=k\}|}{|T\times V|}.$$
Figure \ref{fig:global_degree} shows that the degree distribution is very heterogeneous, which discards the hypothesis of a normal behaviour. In this situation, one may hardly identify values of degree that could be considered anomalous. \\

\noindent A solution is to fit this distribution, and then find values which deviates from the model. Given its heterogeneity, one may think that it is well fitted by a power law distribution $P(k)\propto k^{-\alpha}$ where $\alpha>1$ and $k_{\text{max}}\geqslant k \geqslant k_{\text{min}}>0$. However, we show that this is not the case following the procedure proposed by Virkar \textit{et al.} \cite{virkar2014power}. Results show that differences between the empirical distribution and the estimated model cannot be attributed to statistical fluctuations, which leads us to reject the hypothesis that the degree is distributed according to a power law distribution.\\

\begin{figure}
\begin{tikzpicture}
\node[] () at (4.8,4) {\resizebox{0.4\textwidth}{!}{\input{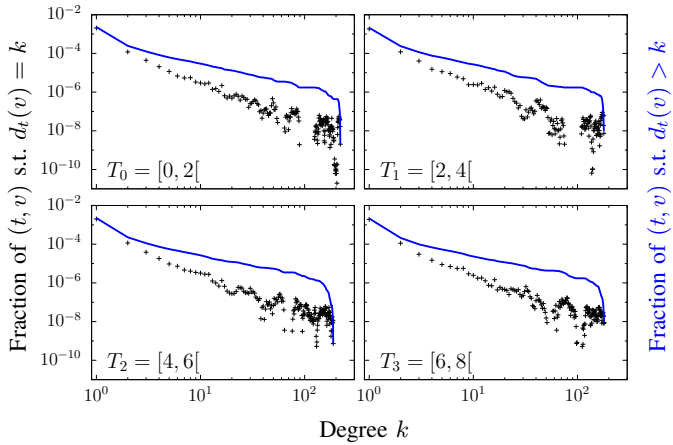}}};
\node[label={[rotate=90]\small{Fraction of $(t,v)$ s.t. $d_t(v)=k$}}] () at (0.5,4){};
\node[label={[rotate=90] \small{\textcolor{blue}{Fraction of $(t,v)$ s.t. $d_t(v)>k$}}}] () at (9,4){};
\node[label={\small{Degree $k$}}] () at (4.8,0.6){};
\end{tikzpicture}
\caption{\small{\textbf{Degree distribution and complementary cumulative degree distribution over 2-second time slices.} For ${T_0=[0,2[}$, $T_1=[2,4[$, $T_2=[4,6[$ and $T_3=[6,8[$, we compute the degree $d_t(v)$ for all $(t,v)$ in the corresponding sub-stream and plot the distribution of the set of values $\mathcal{O}_i=\{d_t(v) \: : \: (t,v) \in T_i \times V\}$, for $i=\{0,1,2,3\}$. The fraction expresses the probability to draw a time instant $t\in T_i$ and a node $v\in V$ such that $d_t(v)=k$.}}
\label{fig:local_degree}
\end{figure}

This shows that finding outliers in this type of distribution is not trivial. In order to circumvent this issue, we observe degrees on sub-streams corresponding to IP traffic during time slices of duration $\tau=2.0\text{s}$. Formally, we call $T_i=[2i,2i+2[$ the $i^{th}$ time slice, for all $i \in \{0, \dots, 1799\}$, and we define
$$f_i(k)=\frac{|\{(t,v)\in T_i \times V \::\: d_t(v)=k\}|}{|T_i\times V|},$$
the degree distribution of the $i^{\text{th}}$ time slice. Figure \ref{fig:local_degree} shows that these distributions still are heterogeneous.\\

\begin{figure}[ht]
\hspace*{0.3cm}
\scalebox{0.6}{
\input{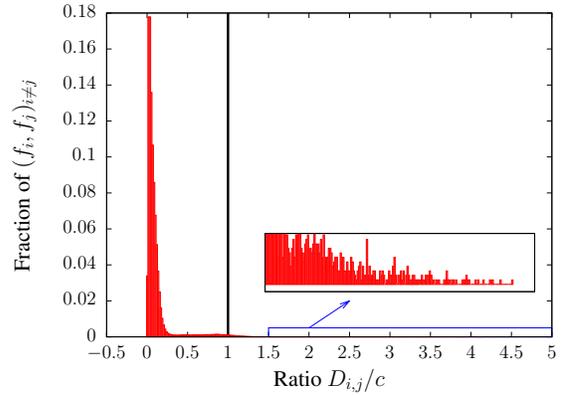}
}
\caption{\small{\textbf{Similarity of degree distributions over 2-seconds time slices.} For all pairs of degree distributions $(f_i,f_j)_{i\neq j}$, we compute the ratio between the KS distance $D_{i,j}$ and the critical value $c$. We plot the distribution of the set of ratios $D_{i,j}/c$ for $i,j\in \{0,\dots,1799\}, i\neq j$. We can see that most values are below 1 meaning that most KS distances are smaller than $c$. Accordingly, based on the two sample KS test, most degree distributions are similarly distributed. On the other hand, $1\%$ of the ratios are larger than 1, which is the result of the comparison between some deviating distributions and all others.}}
\label{fig:two_sample}
\end{figure}

\noindent Nonetheless, Figure \ref{fig:local_degree} also shows that degree distributions $f_i$ have similar shapes. To quantify this similarity, we perform two-sample Kolmogorov-Smirnov (KS) tests on all pairs of distributions $(f_i,f_j)_{i\neq j}$ \cite{Press92numericalrecipes}. According to the relative position between the KS distance $D_{i,j}$ and a critical value $c$, this test assess whether two samples may come from the same distribution or not. Let $m=k_{\text{max}}^i$ and $n=k_{\text{max}}^j$, be the sizes of the two samples. With a significance level of $0.1$, ${c=1.073\sqrt{\frac{n+m}{nm}}}$ \cite{agarwal2006basic}. Figure \ref{fig:two_sample} shows the ratio between $D_{i,j}$ and $c$. Most $D_{i,j}$ are below $c$. This means that most samples $\mathcal{O}_i=\{d_t(v) \: : \: (t,v) \in T_i \times V\}$ are drawn from the same distribution. On the contrary, some of them are different from all others which in turn, indicate changes in the overall behaviour on particular sub-streams.\\

Using these observations, we design below an outlier detection method based on the temporal homogeneity of heterogeneous degree distributions.

\section{Leveraging Temporal Homogeneity to Detect Events}
\label{sec:detection}

The above observations lead to the following conclusion: degree distributions are heterogeneous \textit{in the same way} on most, if not all, time slices. In other words, in each time slice, the fraction of couples $(t, v)$ that have a given degree is similar to this fraction in other time slices. This is what we will consider as \textit{normal}. Anomalies, instead, correspond to significant deviation from the usual fraction of nodes having a given degree. In this section we describe our method to compare degree distributions on all time slices and its use for outlier detection.\\

\begin{figure*}
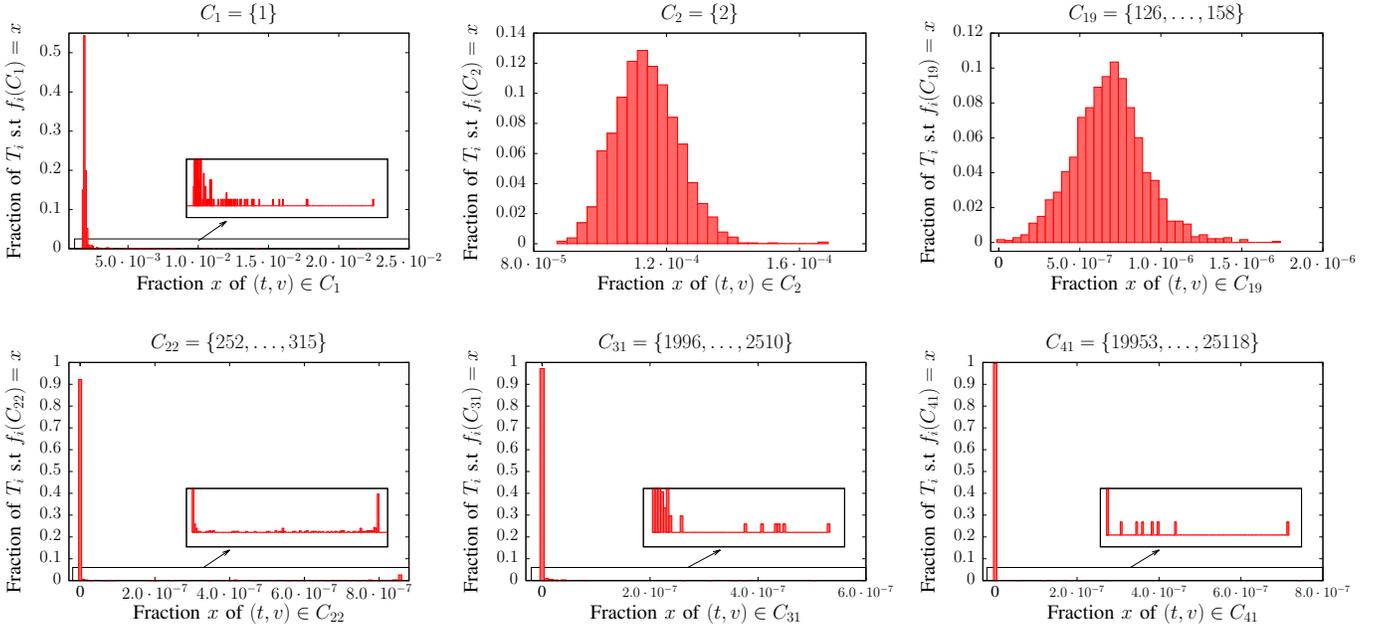

	\begin{subfigure}[t]{0.33\textwidth}
	  \resizebox{0.95\textwidth}{!}{\input{M_1_hist_first_removal.tex}}
	\end{subfigure}
	\begin{subfigure}[t]{0.33\textwidth}
	    \resizebox{0.95\textwidth}{!}{\input{M_4_hist_first_removal.tex}}
	    	\end{subfigure}
	\begin{subfigure}[t]{0.33\textwidth}
	    \resizebox{0.95\textwidth}{!}{\input{M_22_hist_first_removal.tex}}
	    	\end{subfigure}

\vspace*{0.4cm}
	\begin{subfigure}[t]{0.33\textwidth}
	  	    \resizebox{0.95\textwidth}{!}{\input{M_25_hist_first_removal.tex}}	 
	    	\end{subfigure}	    	
	\begin{subfigure}[t]{0.33\textwidth}
		    \resizebox{0.95\textwidth}{!}{\input{M_34_hist_first_removal.tex}}
	    	\end{subfigure}
	\begin{subfigure}[t]{0.33\textwidth}
		    \resizebox{0.95\textwidth}{!}{\input{M_44_hist_first_removal.tex}}
	    	\end{subfigure}

\vspace*{0.4cm}	    		    	
	    	\caption{\small{\textbf{Distributions of fractions $f_i(C)$ on all time slices $T_i$ for degree class $C$ in $\{C_1, C_2, C_{19}, C_{22}, C_{31}, C_{41}\}$ -} Distributions on $C_1$, $C_2$ and $C_{19}$ are homogeneous with outliers. These classes are labelled as $AN$-classes. Distributions on $C_{22}$, $C_{31}$ and $C_{41}$ are peaked on zero since in most time slices there are no couple $(t,v)$ in the corresponding class. They are labelled as $A$-classes.}}
	    	\label{fig:result_distrib_homogene}
\end{figure*} 
First, notice that it makes little sense to consider the fraction of couples $(t,v)$ having a degree exactly $k$ when $k$ is large: having degree $k-1$ or $k+1$ makes no significant difference. Therefore, to increase the likelihood of observing values in the tail of the distribution, we define logarithmic degree classes $C_j$ and consider the fraction of couples $(t,v)$ having degrees in $C_j$, for all $j$:
$$f_i(C_j)=\frac{|\{(t,v)\in T_i \times V : \: d_t(v)\in C_j\}|}{|T_i\times V|}.$$
We define the $j^{th}$ degree class, $C_j=\{\lceil k_j \rceil, \dots, \lfloor k_{j+1} \rfloor -1 \}$ such that $k_1=1$ and $\log (k_{j+1})=\log (k_j)+r$ where $r=0.1$ is the degree class size. This leads to $C_1=\{1\}$, $C_2=\{2\}$, $C_3=\{3\}$, $C_4=\{4,5\}$, etc., until $C_{41}=\{19953,\dots,25117\}$. Then, to compare degree distributions, we plot for a given degree class $C$, the distribution on all time slices $T_i$ of the fraction $f_i(C)$. In other words, we study how the fraction of couples $(t,v)$ which have a degree within $C$ during $T_i$ is distributed among all time slices.\\

Figure~\ref{fig:result_distrib_homogene} shows the distributions for classes $C_1$, $C_2$, $C_{19}$, $C_{22}$, $C_{31}$ and $C_{41}$. In accordance with temporal homogeneity, we can see that most fractions are distributed around the mean and that only a few are distant from it. As expected according to the heterogeneity of degrees, the higher the degree class, the lower the fraction of couples $(t,v)$ within the class. We see on $C_1$ that the average fraction over all time intervals is $2.1\cdot 10^{-3}$. When switching to $C_2$, it drops to $1.15\cdot 10^{-4}$ and gradually decreases to reach $0$ in classes of degrees above $252$. In these high degree classes, the spike on fraction $0$ indicates that in most time slices, there is no couple $(t, v)$ reaching such high degrees. Note that this is a peculiarity of this dataset, one or multiple nodes could have a constant high degree and lead to a nonzero average fraction (see Section \ref{sec:datasets}). In the following, we will refer to these classes as $A$-classes since they only contain abnormal traffic. In opposition, classes having an average fraction greater than 0, which contain abnormal traffic and normal traffic, will be referred to as $AN$-classes.\footnote{In practice, we do not observe homogeneous classes without outliers, hence the lack of $N$-classes.}\\

\begin{figure}
\begin{subfigure}[t]{0.38\textwidth}
\resizebox{0.95\textwidth}{!}{\input{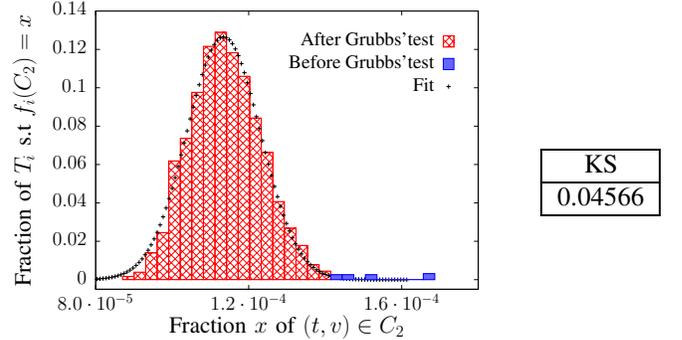}}
\end{subfigure}
\begin{subfigure}[t]{0.1\textwidth}
\vspace*{-2.5cm}
\begin{tabular}{|l|c|r|}
  \hline
  $\:\:\:$ KS \\
  \hline
  0.04566 \\
  \hline
\end{tabular}
\end{subfigure}
\caption{\small{\textbf{Fit of fractions distribution on $C_2$ after removing outliers with Grubbs test -} The KS distance between the fit and the empirical distribution is below the critical value. Hence, this distribution is flagged as an homogeneous distribution with outliers ($AN$-class).}}
\label{fig:fit_grubb}
\end{figure}

In order to validate fractions $f_i$ homogeneity over time slices within each degree class, we fit their distributions with a normal distribution model $P(x)=\frac{1}{\sqrt{2\pi \sigma^2}}e^{-\frac{1}{2}(\frac{x-\mu}{\sigma})^2}$ where values are normally distributed around a mean $\mu$ with a standard deviation $\sigma$. Deciding whether a given distribution is homogeneous with outliers or not may be done as follows~\cite{latapy2013detecting}: (1) Iteratively remove outliers from the distribution with Grubbs test \cite{grubbs1969procedures}; (2) Fitting the resulting distribution with the normal model; (3) Evaluate the goodness of the fit. We use Maximum Likelihood Estimation (MLE) to determine which model parameters fit the best the empirical distribution and evaluate the goodness of the fit with the KS test between empirical and estimated distributions. In this framework, we find 37 distributions that are homogeneous with outliers among the 41 corresponding to each degree class (see Figure \ref{fig:fit_grubb}). The remaining 4 classes are discarded from the study, we call them $R$-classes for rejected classes. One may use more complex and accurate techniques to automatically perform the fit, see for instance the work performed by Motulsky \textit{et al.} \cite{motulsky2006detecting}.\\

Given an homogeneous distribution with outliers, we use here the classical assumption that a value is anomalous if its distance to the mean exceeds three times the standard deviation \cite{chandola2009anomaly}, \cite{han2011data}. In classes displayed in Figure \ref{fig:result_distrib_homogene}, we obtain 151 time slices flagged as anomalous in the first class containing degree 1 only, 5 anomalous time slices in $C_2$ and 12 in $C_{19}$. In $A$-classes, peaked on 0, anomalous values correspond to all values greater than 0.\\

All in all, our method for event detection from degree distributions is the following: we group degree values into degree classes of logarithmic width. For a given degree class $C$, we look at the distribution on all time slices of the fraction $f_i(C)$. This distribution indicates anomalous values which means that there are anomalous high numbers of couples $(t,v)$ having degree within $C$ during specific time slices $T_i$. We then call an anomalous value of this kind a detected event and denote it $(C,\,T_i)$.\\
 
A detected event gives two pieces of information: the time slice $T_i$ on which the anomalous value has been observed, and the degree class $C$ in which the couples responsible for the high fraction are located. At this stage, we detected 1,358 such events. However, a time slice and a degree class are not sufficient information to accurately characterize the anomaly. We now address the goal of identifying couples $(t,v)$ in $T\times V$ responsible for these detected events. 


\section{Iterative Removal to Identify Events}
\label{sec:identification}

A detected event $(C,\,T_i)$ is a degree class $C$ and a time slice $T_i$ such that the fraction $f_i(C)$ is unusually high compared to the ones in other time slices. Identifying this event means recovering the set $\mathcal{I}_{(C,\,T_i)}$ of couples $(t,v)$ responsible for this anomaly. In this section, we introduce an iterative removal method and show that it leads to such identification.\\

\begin{figure}
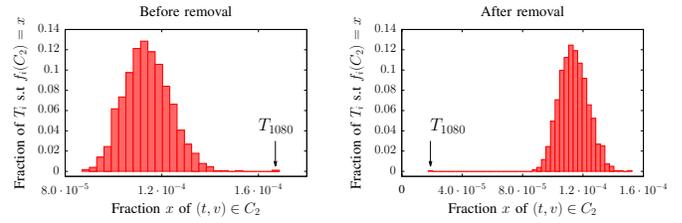

	\begin{subfigure}[t]{0.24\textwidth}
	  \resizebox{0.95\textwidth}{!}{\input{removal_outlier_low_class2_before.tex}}
	\end{subfigure}
	\begin{subfigure}[t]{0.24\textwidth}
	    \resizebox{0.95\textwidth}{!}{\input{removal_outlier_low_class2_after.tex}}
	    	\end{subfigure}
\caption{\small{\textbf{Incorrect identification in $AN$-classes -} The removal of all interactions $(t,uv)$ such that $d_t(v)$ is in $C_2$ during the detected time slice $T_{1080}$ causes the appearance of a negative outlier. Note that the resulting fraction on $T_{1080}$ is not zero since the removal of some interactions has decreased the degree of nodes in higher classes which end up having a degree in $C_2$.}}
	    	\label{fig:removal_false_identification}
\end{figure}

Let us take event $(C_2,T_{1080})$ as an example. We have access to the set of couples $(t,v)$ which have a degree in $C_2$ during $T_{1080}$. However, we cannot directly identify the event by this set. Indeed, let us consider the new link stream $L'$ in which we removed the corresponding interactions: $L'=(T,V,E')$ with $E'=E\setminus \{(t,uv) : t\in T_{1080}\, \text{ and } \, d_t(v)\in C_2 \}$. We see in Figure \ref{fig:removal_false_identification} that the removal of this set of interactions from the link stream causes the appearance of a negative outlier\footnote{We call negative outlier an outlier which is lower than the mean.} in the distribution of fractions on $C_2$. Thus, by removing all interactions $(t,uv)$ such that couples $(t,v)$ have degree in $C_2$ during $T_{1080}$, we removed anomalous traffic but also normal traffic. Therefore, identifying the detected event $(C_2,\,T_{1080})$ as the set $\mathcal{I}_{(C_2,\,T_{1080})}=\{(t,v) : t\in T_{1080}\, \text{ and } \, d_t(v)\in C_2\}$ is not accurate enough.\\

This suggests that one cannot directly identify couples acting abnormally in $AN$-classes. Indeed, in these classes, the normal fraction is greater than zero. Hence, an anomalous fraction consists in anomalous couples but also normal ones, which prevents us from identifying responsible couples without disrupting normal traffic.\\

\begin{figure}
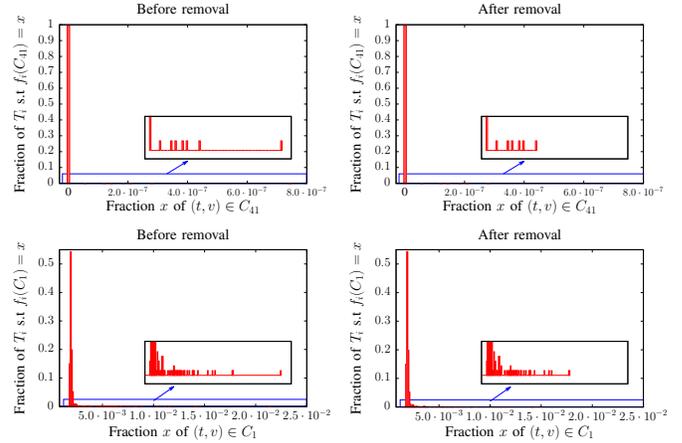

	\begin{subfigure}[t]{0.24\textwidth}
	  \resizebox{0.95\textwidth}{!}{\input{M_44_hist_0_removal_ex.tex}}
	\end{subfigure}
	\begin{subfigure}[t]{0.24\textwidth}
	    \resizebox{0.95\textwidth}{!}{\input{M_44_hist_1_removal_ex.tex}}
	    	\end{subfigure}
	
	\vspace*{0.1cm}    	
	\begin{subfigure}[t]{0.24\textwidth}
	  \resizebox{0.95\textwidth}{!}{\input{M_1_hist_0_removal_ex.tex}}
	\end{subfigure}
	\begin{subfigure}[t]{0.24\textwidth}
	    \resizebox{0.95\textwidth}{!}{\input{M_1_hist_1_removal_ex.tex}}
	    	    	\end{subfigure}	    	
	    	    	
\caption{\small{\textbf{Event identification in $A$-classes -} The removal of an identified event in the $A$-class $C_{41}$ allows the identification of an event detected in the $AN$-class $C_1$.}}
	    	\label{fig:removal_41}
\end{figure}

On the contrary, in $A$-classes the expected fraction is zero. Therefore, couples $(t,v)$ contributing to non-zero fractions are clearly anomalous. Events detected in such degree class $C$ can therefore be correctly identified with the set $\mathcal{I}_{(C,\,T_i)}=\{(t,v) : t\in T_{i}\, \text{ and } \, d_t(v)\in C\}$. Thus, we now consider A class $C_{41}$. Its larger anomalous fraction corresponds to time slice $T_{315}$. Hence, this event can be identified by the set $\mathcal{I}_{(C_{41},\,T_{315})}=\{(t,v) : t\in T_{315}\, \text{ and } \, d_t(v)\in C_{41}\}$. Figure~\ref{fig:removal_41} shows the consequences of the removal of these abnormal couples activities. As expected, the anomalous fraction in $C_{41}$ vanishes without creating a negative outlier. Additionally, we notice the disappearance of event $(C_1,T_{315})$. Indeed, removed nodes $v$ were linked to an unexpectedly large number of nodes having degree 1 before the removal. Then, the removal of the set $\mathcal{I}_{(C_{41},\,T_{315})}$ leads to the identification of event $(C_1,\,T_{315})$ such that $\mathcal{I}_{(C_1,\,T_{315})}=\{(t,u) : t\in T_{315}, u\in N_t(v), d_t(v)\in C_{41} \, \text{ and } \, d_t(u)\in C_1\}$, where $N_t(v)$ is the set of neighbours of $v$ at time $t$.\\

All in all, our approach for event identification is the following. For each detected and identified event $(C,\,T_i)$ in $A$-classes, we remove abnormal activities of couples $(t,v)\in \mathcal{I}_{(C,\,T_i)}$ such that on the $n^{\text{th}}$ removal, we consider the link stream $L_n(V,T,E_n)$ with $E_n=E_{n-1}\setminus \{(t,uv) : t\in T_i\, \text{ and } \, d_t(v)\in C\}$ and $E_0=E$. In addition to removing anomalous traffic identified in $A$-classes, this process allows to identify related events in $AN$-classes as well. If a given removal creates a negative outlier in a degree class, this means that we removed too much. The removal that caused it is then cancelled and the corresponding event stays detected but unidentified.\\

\begin{figure}
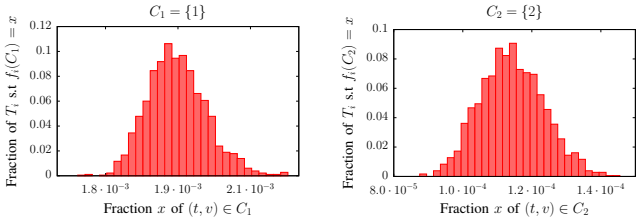

	\begin{subfigure}[t]{0.24\textwidth}
	  \resizebox{0.95\textwidth}{!}{\input{M_1_hist_final_removal.tex}}
	\end{subfigure}
	\begin{subfigure}[t]{0.24\textwidth}
	    \resizebox{0.95\textwidth}{!}{\input{M_4_hist_final_removal.tex}}
	    	\end{subfigure}
\caption{\small{\textbf{Distributions of fractions on all time slices for degree classes $C_1$ and $C_2$ after event removals -} Before event removal there were 151 anomalous values in $C_1$ and 5 in $C_2$. After removal, it only remains 10 unidentified anomalous values in $C_1$ and 2 in $C_2$.}}
	    	\label{fig:final_distrib}
\end{figure}

\begin{figure}
\hspace*{-0.3cm}
\begin{tikzpicture}
\node[] () at (4.8,4) {\resizebox{0.4\textwidth}{!}{\input{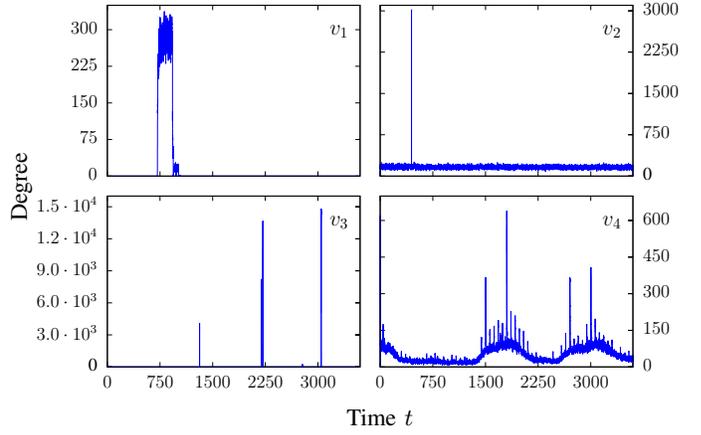}}};
\node[label={[rotate=90]\small{Degree}}] () at (0.3,4){};
\node[label={\small{Time $t$}}] () at (4.8,0.7){};
\end{tikzpicture}

\caption{\small{\textbf{Degree profiles of four identified nodes - }$v_1$ is responsible for outliers observed in $C_{22}=\{252,...,2510\}$. The set $\{(t,v_1) : t\in [712,940[ \, \text{ and } \, d_t(v)\in C_{22}\}$ has been identified and removed. $v_2$ has a normal activity with a degree around 160 and a sharp variation on $T_{223} = [446,448[$. The set $\{(t,v_2) : t\in T_{223} \, \text{ and } \, d_t(v)\in C_{32}\}$ has been identified and removed. Sets $\{(t,v_3)\}$ where $v_3$ is active have all been identified and removed. For node $v_4$, the four peaks corresponding to  degree values higher than 300 have been identified and removed. The degree profiles of these nodes suggest that they constitute malicious activity \cite{huang2014network,mazel2014taxonomy}. Node $v_3$ reaches several powers of two indicating that it is running network scans. We observe a similar behaviour around a degree of $256$ for node $v_1$.}}
	    	\label{fig:profil_degre}
\end{figure}

In our dataset, none of the removals generated negative outliers (this is not what we observe for all datasets, see Section \ref{sec:datasets}). Altogether, we directly identified and removed 205 events in $A$-classes. These removals allowed us to identify a total of $1,163$ outliers on the $1,358$ previously detected ones, hence more than $85\%$ of detected outliers. To do so, we removed $7.4\%$ of all the traffic. We can see in Figure \ref{fig:final_distrib} the final shape of classes $C_1$ and $C_2$ in which almost all outliers disappeared. Figure \ref{fig:profil_degre} shows the degree profiles of 4 nodes which have been removed for time periods during which they were acting abnormally.  \\

\section{Validation}
\label{sec:Validation} 

The IP traffic trace we use does not have a ground truth dataset listing abnormal IP addresses and instants. However, we can validate our results by looking at the consequences of the removals on the average degree per second.\\

\begin{figure}
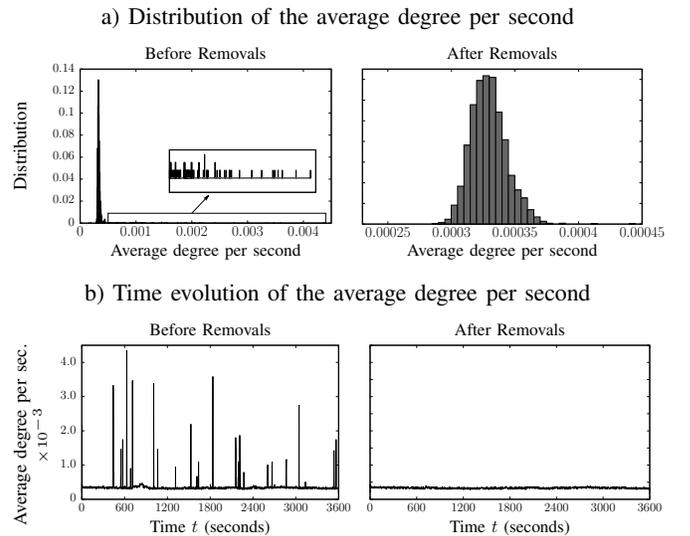

{\center\small{a) Distribution of the average degree per second\\}}

\vspace*{0.15cm}
\begin{subfigure}[t]{0.08\textwidth}
\vspace*{-2.2cm}
\begin{tikzpicture}
\node[label={[rotate=90]\scriptsize{Distribution}}] () at (0.5,0){};
\end{tikzpicture}
	\end{subfigure}\hspace*{-1.1cm}
	\begin{subfigure}[t]{0.24\textwidth}
	  \resizebox{0.95\textwidth}{!}{\input{distrib_degre_moyen_removal_0.tex}}
	\end{subfigure}\hspace*{-0.3cm}
\begin{subfigure}[t]{0.24\textwidth}
	    \resizebox{0.95\textwidth}{!}{\input{distrib_degre_moyen_removal_1.tex}}
	    	\end{subfigure}

{\center\small{b) Time evolution of the average degree per second\\	}}

\vspace*{0.15cm}
	\begin{subfigure}[t]{0.08\textwidth}
\begin{tikzpicture}
\node[label={[rotate=90]\scriptsize{Average degree per sec.}}] () at (0.5,-5){};
\node[label={[rotate=90]\tiny{$\times 10^{-3}$}}] () at (0.75,-5){};
\end{tikzpicture}
	\end{subfigure}\hspace*{-1cm}
	    	\begin{subfigure}[t]{0.24\textwidth}
	    \resizebox{0.95\textwidth}{!}{\input{degre_moyen_before.tex}}
	    	\end{subfigure}\hspace*{-0.3cm}
	    	\begin{subfigure}[t]{0.24\textwidth}
	    \resizebox{0.95\textwidth}{!}{\input{degre_moyen_after.tex}}
	    	\end{subfigure}
\caption{\small{\textbf{Consequences of event removals on the average degree per second - }Our method succeeds in removing identified anomalies with no significant impact on the underlying normal traffic.}}
	    	\label{fig:volume_feature}
\end{figure}

Let $d_t$ be the instantaneous average degree at time $t$: ${d_t=\frac{1}{V} \sum_{v} d_t(v)}$. The average degree during second ${s_i=[i,i+1]}$, denoted by $d(s_i)$, is the average of $d_t$ from $t=i$ to $t=i+1$, for all ${i\in \{0, \cdots, 3599\}}$: $$d(s_i)= \int_i^{i+1} d(t)\, dt \: .$$

We see in Figure \ref{fig:volume_feature}.a that the average degree per second is homogeneously distributed with outliers. After removing events identified with our method, we see in Figure 11.b that peaks as well as sudden changes in the trend disappear, while the average over time stays unaltered. Likewise, in the distribution, all outliers disappear but the bell curve stays the same. Quantitatively, we find $33$ outlying seconds before removals versus $5$ after applying our method. The average over time of the average degree per second is equal to $3.39571\cdot 10^{-4}$ before removals, and to $3.31795\cdot 10^{-4}$ after removals. These results show that our method succeeds in removing abnormal traffic without altering the underlying normal traffic.\\

These results highlight another important fact. An outlier in distribution of Figure \ref{fig:volume_feature}.a means that there is a second during which the average degree is larger than usual. This event is detected but not identified since we cannot trace back responsible nodes with this aggregated feature. By using the instantaneous degree on couples $(t,v)$, our method is able to identify events unidentified with the average degree per second. This last result is particularly promising: it shows that by using more complex and less aggregated features, it is possible to identify events previously detected but unidentified with simpler metrics. Then, the 195 events that we were not able to identify with the instantaneous degree of nodes could therefore be identified in future works by using other link stream features.

\section{Other Datasets}
\label{sec:datasets}
To test the generality and applicability of our method, we test it on other datasets from the MAWI archive. In this section, we present the main results and differences we observe with these datasets.

\subsection{One day long IP traffic trace from June 2013}
We use here a one day long IP traffic capture from the MAWI archive from June $25^{\text{th}}$, 2013, at 00:00 to June $26^{\text{th}}$, 2013, at 00:00. The set $\mathcal{D}$ contains $2,196,079,591$ triplets involving $15,390,238$ different IP addresses. This dataset is larger than the first one and covers one day of IP traffic with its circadian cycle. We keep identical time slices size and degree classes size. \\

\begin{figure}[]
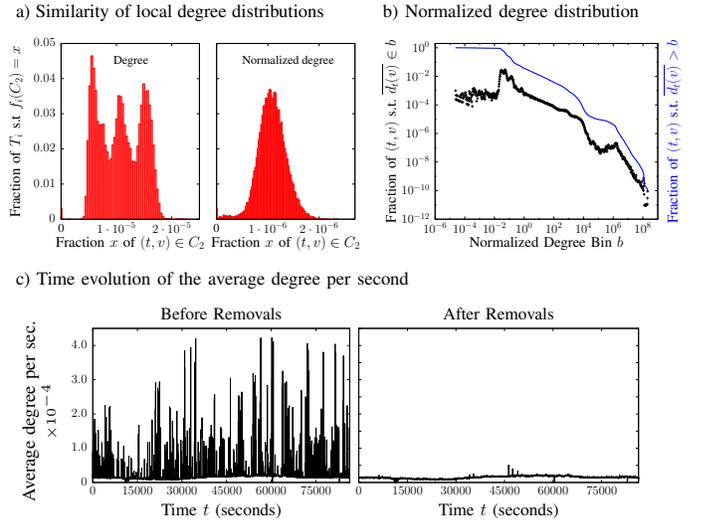

	\begin{subfigure}[t]{0.24\textwidth}
\scriptsize{a) Similarity of local degree distributions}\hspace*{0.25cm}

\vspace*{0.15cm}
\hspace*{0.45cm}\resizebox{0.95\textwidth}{!}{\input{normalized_degree.tex}}
\end{subfigure}\hspace*{0.45cm}
\begin{subfigure}[t]{0.24\textwidth}
\scriptsize{b) Normalized degree distribution}\hspace*{0.25cm}

\vspace*{0.15cm}
\resizebox{0.95\textwidth}{!}{\input{degprofile_4.tex}}
\end{subfigure}

\vspace*{0.3cm}
\begin{subfigure}[t]{0.49\textwidth}
\scriptsize{c) Time evolution of the average degree per second \hspace*{1cm}}

\vspace*{0.15cm}
\begin{subfigure}[t]{0.05\textwidth}
\vspace*{-2.7cm}
\begin{tikzpicture}
\node[label={[rotate=90]\scriptsize{Average degree per sec.}}] () at (0.5,-5){};
\node[label={[rotate=90]\tiny{$\times 10^{-4}$}}] () at (0.75,-5){};
\end{tikzpicture}
	\end{subfigure}
\begin{subfigure}[t]{0.49\textwidth}
\resizebox{0.95\textwidth}{!}{\input{degre_moyen_before_one_day.tex}}
\end{subfigure}\hspace*{-0.6cm}
\begin{subfigure}[t]{0.49\textwidth}
\resizebox{0.95\textwidth}{!}{\input{degre_moyen_after_one_day.tex}}
\end{subfigure}
\end{subfigure}
\caption{\small{\textbf{Results on a one day long IP traffic trace - } After degree normalization, both the similarity of local distributions and the heterogeneity of the global distribution are verified. Results are validated by the average degree per second after removals in which the normal traffic is preserved. Note that in the global distribution (b), the normalized degree has been rescaled with a constant to have a better range of values and values are merged into bins to smooth the distribution.}}
\label{fig:oneday}
\end{figure}
Figure \ref{fig:oneday}.a (left) shows, for $C_2=\{2\}$, the distribution on all time slices $T_i$ of the fraction $f_i(C_2)$. Partly as a result of circadian cycles, we see that this distribution consists of three normal distributions. To address this issue, we normalize the degree with the average degree per second and consider the normalized degree, denoted by $\overline{d_t(v)}$, such that 
$$\overline{d_t(v)}=\frac{d_t(v)}{d(\lfloor t \rfloor)}\:, $$
where $\lfloor t \rfloor$, the floor function of $t$, is the second to which $t$ belongs. We see in Figure \ref{fig:oneday}.a (right) that local distributions on time slices are similar and in Figure \ref{fig:oneday}.b that the global normalized degree distribution is heterogeneous. Thus, the two constraints required to apply our method are met. We find 34 degree classes, from $C_1=\{1\}$ to $C_{34}=\{3982,\dots,5011\}$. Among these, $3$ classes are rejected because they do not fit with an homogeneous distribution with outliers. We find $11$ $AN$-classes from $k=1$ to $k=26$ and $20$ $A$-classes from $k=51$ to $k=5011$. We detect $22,669$ outliers and succeed in identifying $63\%$ of them. To do so, we removed $8.7\%$ of all the traffic. Once again, we see in Figure \ref{fig:oneday}.c that these removals lead to the cleaning of the average degree per second: $1,672$ abnormal seconds before removals and $10$ after. Likewise, normal traffic stays unchanged: the average over time of the average degree per second is equal to $1.53199\cdot 10^{-5}$ before removals and to $1.48824\cdot 10^{-5}$ after.

\subsection{Fifteen-minute IP traffic trace from November 2018: comparison to MAWILab}
We use here a fifteen-minute IP traffic capture from the MAWI archive on November $3^{\text{rd}}$, 2018, from 14:00 to 14:15. The set $\mathcal{D}$ contains $64,913,871$ triplets involving $16,453,608$ different IP addresses. This dataset is more recent than the first one and has a list of anomalies indexed by MAWILab \cite{fontugne2010mawilab} to which we can compare our results. Given the shorter temporal extent, we take time slices of size $\tau=1.0\text{s}$ instead of $\tau=2.0\text{s}$, in order to keep a significant number of time slices. Degree classes stay unchanged. \\

\begin{figure*}[]
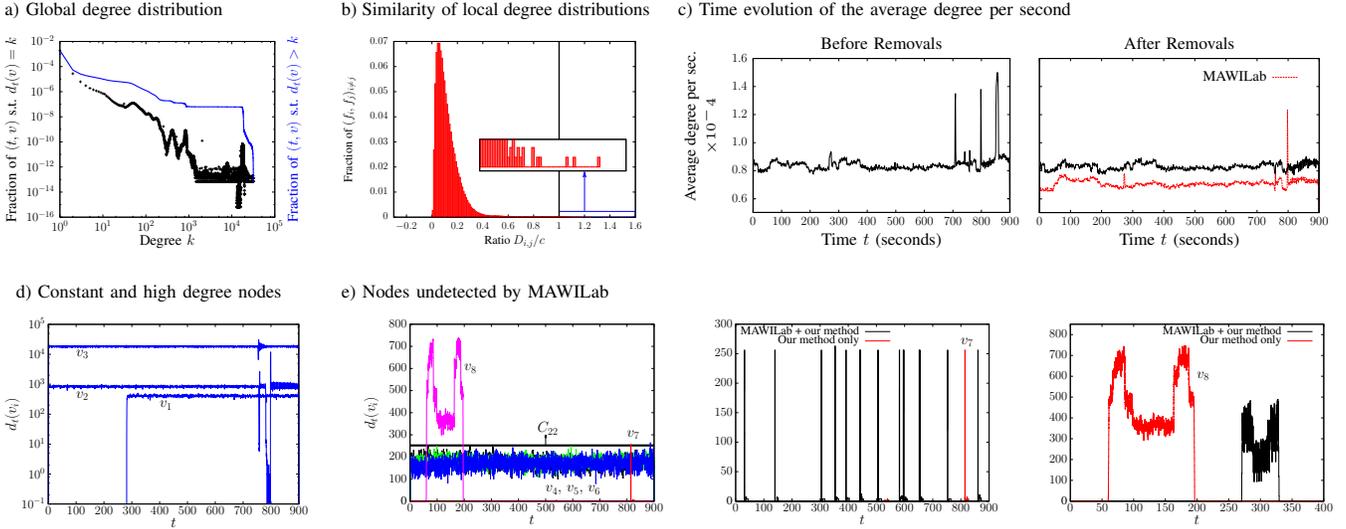

	\begin{subfigure}[t]{0.24\textwidth}
\scriptsize{a) Global degree distribution}\hspace*{0.25cm}

\vspace*{0.15cm}
\resizebox{0.95\textwidth}{!}{\input{degprofile_3.tex}}
\end{subfigure}
\begin{subfigure}[t]{0.24\textwidth}
\scriptsize{b) Similarity of local degree distributions}\hspace*{0.25cm}

\vspace*{0.15cm}
\resizebox{0.95\textwidth}{!}{\input{two_sample_ks_one_quarter.tex}}
\end{subfigure}
\begin{subfigure}[t]{0.49\textwidth}
\scriptsize{c) Time evolution of the average degree per second \hspace*{1cm}}

\vspace*{0.15cm}
\begin{subfigure}[t]{0.05\textwidth}
\vspace*{-2.7cm}
\begin{tikzpicture}
\node[label={[rotate=90]{\fontsize{6}{6}\selectfont Average degree per sec.}}] () at (0.5,-5){};
\node[label={[rotate=90]\tiny{$\times 10^-4$}}] () at (0.75,-5){};
\end{tikzpicture}
	\end{subfigure}
\begin{subfigure}[t]{0.49\textwidth}
\resizebox{0.95\textwidth}{!}{\input{degre_moyen_before_one_quarter.tex}}
\end{subfigure}\hspace*{-0.3cm}
\begin{subfigure}[t]{0.49\textwidth}
\resizebox{0.95\textwidth}{!}{\input{degre_moyen_after_one_quarter.tex}}
\end{subfigure}
\end{subfigure}

\vspace*{0.5cm}
	\begin{subfigure}[t]{0.24\textwidth}
\hspace*{0.15cm}\scriptsize{d) Constant and high degree nodes}\hspace*{0.25cm}

\vspace*{0.15cm}
\resizebox{0.95\textwidth}{!}{\input{profil_N_classes_degre_eleve_moy.tex}}
\end{subfigure}
\begin{subfigure}[t]{0.74\textwidth}
\scriptsize{e) Nodes undetected by MAWILab}\hspace*{0.25cm}

\vspace*{0.15cm}
\hspace*{0.3cm}\begin{subfigure}[t]{0.32\textwidth}
\resizebox{0.95\textwidth}{!}{\input{profil_pas_dans_mawilab_moy.tex}}
\end{subfigure}\hspace*{0.1cm}
\begin{subfigure}[t]{0.32\textwidth}
\resizebox{0.95\textwidth}{!}{\input{erreur_mawilab.tex}}
\end{subfigure}\hspace*{0.1cm}
\begin{subfigure}[t]{0.32\textwidth}
\resizebox{0.95\textwidth}{!}{\input{erreur2_mawilab.tex}}
\end{subfigure}
\end{subfigure}

\caption{\small{\textbf{Fifteen minutes from November the $3^{\text{rd}}$ $2018$ - } (a - b) The heterogeneity of the global degree distribution and the similarity of local degree distributions are verified. (c) Results are validated by the average degree per second after removals. Contrary to our method, we observe a decrease of the average over time when removing events identified by MAWILab. (d) $AN$-classes are observed in high degree classes: $v_1$ is responsible for the normal traffic observed in $C_{24}=\{399,\dots,502\}$; $v_2$ for the one in $C_{27}=\{795,\dots,1001\}$ and $v_3$ for the one in $C_{40}=\{15849,\dots,19953\}$. (e) Nodes $v_2$, $v_4$, $v_5$, $v_6$, $v_7$ and $v_8$ are not detected by MAWILab. However, we see that the removal of the abnormal activity of $v_2$ is responsible for the disappearance of the spike around $t=800$ in (c) and that $v_7$ and $v_8$ have a suspicious activity which is usually detected by MAWILab. Note that high degree node $v_3$ in (d) has been removed from the calculation in the average degree per second in (c) to have a greater clarity.}}
\label{fig:onequarter}
\end{figure*}

We observe an heterogeneous global degree distribution and similar local degree distributions on time slices of $1.0$ seconds (see Figures \ref{fig:onequarter}.a and \ref{fig:onequarter}.b). We find $43$ degree classes, from $C_1=\{1\}$ to $C_{43}=\{31623,\dots,39810\}$. Among these, there are $23$ $AN$-classes, $17$ $A$-classes and $3$ $R$-classes. \\

Contrarily to previous datasets, we observe three $AN$-classes in high degree classes: $C_{24}=\{399,\dots,502\}$, $C_{27}=\{795,\dots,1001\}$ and $C_{40}=\{15849,\dots,19953\}$. We can see in Figure \ref{fig:onequarter}.d, that this is due to three nodes which have a constant degree fluctuating within each of these classes and, as a result, form the observed normal traffic.\\

In this dataset, several removals generated negative outliers. For instance, the removal of event $(C_{40},T_{752})$ generated a negative outlier in class $C_1$. The corresponding event was incorrectly identified by the set $\mathcal{I}_{(C_{40},\,T_{752})}=\{(t,v) : t\in T_{752}\, \text{ and } \, d_t(v)\in C_{40}\}$. Indeed, this event corresponds to the spike of activity of node $v_3$ from $755.3$ to $756.5$ (see Figure \ref{fig:onequarter}.d). Yet, the normal behaviour of $v_3$ is to be linked to an average of $18,178$ nodes of degree $1$ over time. Thus, the removal of its activity during this time period leads to a negative outlier in $C_1$, since, in addition to removing abnormal interactions of $v_3$, it also removes its legitimate interactions. This shows that to identify event $(C_{40},T_{752})$, we need to use a finer and more complex feature than the degree. \\

Finally, our method enabled us to detect $827$ outliers and identify $796$ of them ($96\%$). To do so, we removed $1.2\%$ of all the traffic. We see in Figure \ref{fig:onequarter}.c that, as with the two other datasets, removals lead to a traffic free of most degree-related anomalies. The number of abnormal seconds is equal to $19$ before removals versus $0$ after removals. Likewise, the average over time of the average degree per second goes from $9.311934\cdot 10^{-5}$ to $9.198076\cdot 10^{-5}$ after removals.\\

This dataset contains a MAWILab database to which we can compare our results \cite{fontugne2010mawilab}. It lists and labels anomalies in traffic from the MAWI archive by using a graph-based methodology that compares and combines the output of several independent anomaly detectors. On November $3^{\text{rd}}$, 2018, from 14:00 to 14:15, it indicates a total of $287$ anomalous IP addresses. To each of these is associated a time period during which it is evaluated as abnormal and a label classifying its anomaly type among the following categories~\cite{mazel2014taxonomy}:\\
-- Point to point denial of service: a large number of packets are sent between two IP addresses;\\
-- Distributed denial of service: a large number of packets are sent between multiple sources and one destination;\\
-- Network scan: an IP address scans a network of several destination IP addresses;\\
-- Port scan: an IP address scans several ports of one destination;\\
-- Point multipoint: normal router traffic;\\
-- Alpha flow: normal peer to peer traffic;\\ 
-- Other: normal outage traffic;\\
Since we do not consider the port number, and given that the degree feature does not account for the number of exchanged packets, anomalies within the point to point denial of service, port scan and alpha flows categories cannot be detected by our method. Moreover, we do not consider events corresponding to legitimate traffic. This reduces the number of identified IP addresses to $77$.  \\       

With our method, we find $33$ anomalous IP addresses. Six of them are not listed by MAWILab. They correspond to node $v_2$ in Figure \ref{fig:onequarter}.d and nodes $v_4$, $v_5$, $v_6$, $v_7$ and $v_8$ in Figure \ref{fig:onequarter}.e (left). Node $v_2$ has been removed during its spike of activity from $798.18$ to $799.87$. Likewise, node $v_7$ has been removed from $814.06$ to $815.00$ and node $v_8$ on the whole time period during which it is active. Note that, as we can see in Figure \ref{fig:onequarter}.e (middle and right), nodes $v_7$ and $v_8$ activities are typical of nodes performing network scans which are usually detected by MAWILab. The remaining three nodes $v_4$, $v_5$ and $v_6$, have been removed on periods of respectively $0.0768$s, $0.0677$s and $0.181$s because of their ephemeral activity within the $A$-class $C_{22}=\{252,\dots,317\}$. This could be avoided by using larger classes (see section~\ref{sec:construct_class}). \\ 

In the network scan category, we identified $24$ IP addresses among the $76$ ($32.6\%$) listed by MAWILab. All network scans involving more than $250$ different destinations have been identified with our method. As mentioned above, we identified in addition two IP addresses that the MAWILab detectors missed (see Figure \ref{fig:onequarter}.e). Moreover, for the corresponding events, our temporal precision is much better than the one provided by MAWILab detectors. However, our method fails to identify IP addresses permanently linked to the network since they have constant degree profiles and therefore lead to $AN$-classes. More generally, we did not find IP addresses which scan networks involving less than $250$ destinations since all classes below $C_{22}=\{252,\dots,317\}$ are $AN$-classes, and since their activities are not linked to the ones of removed events. Nonetheless, time slices during which most of these network scans occur have been detected as outliers in their corresponding degree class. This inability to identify low degree classes events could be avoided by using a feature different from the degree, in which the corresponding malicious activities deviate more significantly.\\

In the distributed denial of service category, only one anomaly is identified by MAWILab. The corresponding node have a maximum degree of $53$. Hence, we do not find it with our method for the same reasons as above. \\

The six remaining nodes we identified fall in the point to multipoint category that we do not consider as it constitutes normal router traffic.

Finally, Figure \ref{fig:onequarter}.c (right) shows the average degree per second after removing events identified with our method as well as the ones identified in the MAWILab dataset. We see that, with MAWILab, the average over time is affected by the removals. This is mostly due to the poor time precision used by MAWILab to describe anomalies. Indeed, $63\%$ of IP addresses are identified as abnormal on the whole trace, including IP addresses which have a global constant degree with only a few spikes. With our method, instead, when nodes have a constant degree, classes within which their degree fluctuates are labelled as $AN$-classes. Thus, their activities are not removed and normal traffic stays unchanged.

\section{Influence of Parameters}
\label{sec:para}
We showed the efficiency of our method on several datasets. In this section, we perform a series of experiments on the first dataset to study the influence of parameters $\tau$, for time slice duration, and $r$, for degree class size.

\subsection{Variation of Time Slice Sizes}
\label{sec:size_time}

We divide the link stream into time slices of size $\tau$ in order to compare local degree distributions. The success of our method is based on the fact that they are similar from one time slice to another. Due to aggregation over a larger period, it is expected that the larger $\tau$, the larger the similarity between time slices, and conversely when the size decreases.\\

\begin{figure*}
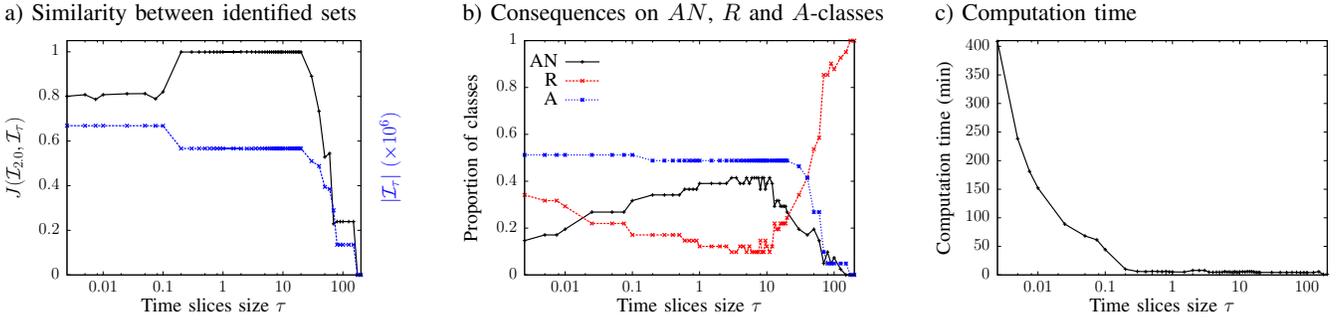

	\begin{subfigure}[t]{0.32\textwidth}
	\small{a) Similarity between identified sets}

	  \resizebox{0.95\textwidth}{!}{\input{jaccard_time_slice.tex}}
	\end{subfigure}\hspace*{0.2cm}
		\begin{subfigure}[t]{0.32\textwidth}
			\small{b) Consequences on $AN$, $R$ and $A$-classes}

	  \resizebox{0.95\textwidth}{!}{\input{proportions_AN_R_A_time_slice.tex}}
	\end{subfigure}\hspace*{0.4cm}
	\begin{subfigure}[t]{0.32\textwidth}
				\small{c) Computation time}

	    \resizebox{0.95\textwidth}{!}{\input{time_time_slice.tex}}
	    	    	\end{subfigure}	  
\caption{\small{\textbf{Time slice size influence -} (a) The Jaccard index between identified sets shows that our method is stable from $\tau=0.2$ to $\tau=20.0$. (b) Small time slices lead to an increase of the number of $A$ and $R$-classes while large time slices lead to a high number of $R$-classes but small number of $AN$ and $A$-classes. (c) The computation time significantly increases as $\tau$ decreases. }}
\label{fig:ts}
\end{figure*}

Let $\mathcal{I}_{\tau}= \cup_{i,j} \, \mathcal{I}_{(C_j,T_i)}$ be the set of identified outliers using time slices of size $\tau$. In order to evaluate the impact of $\tau$, we measure the Jaccard similarity coefficient between $\mathcal{I}_{2.0}$, obtained in the first experiment, and other sets obtained by varying $\tau$:
$$J(\mathcal{I}_{2.0}, \mathcal{I}_{\tau}) = \frac{|\mathcal{I}_{2.0} \cap \mathcal{I}_{\tau}|}{|\mathcal{I}_{2.0} \cup \mathcal{I}_{\tau}|} \; .$$
Results are depicted in Figure \ref{fig:ts}.a. We see that identified sets $\mathcal{I}_\tau$ are identical from $\tau=0.2$ up to $\tau=20.0$. This shows that our method is stable with respect to this parameter. Below this range, we are able to identify slightly more outliers. On the contrary, when the size increases, we identify less and less outliers until no more is identified after $\tau=175.0$. This is explained by the number of $AN$, $A$ and $R$-classes according to $\tau$ in Figure \ref{fig:ts}.b: the more $\tau$ increases, the higher the number of $R$-classes and the lower the number of $A$-classes in which we are able to identify events. When we reach $\tau=175.0$, all classes are rejected, hence no outlier is detected. This increase in the number of rejected classes is provoked by the very small number of time slices when $\tau$ gets larger. Indeed, time slices are insufficiently numerous to establish a normal behaviour and, for all classes $C$, fits between fractions $f_i(C)$ and a normal distribution are more likely to be rejected.  \\

We identify more abnormal couples $(t,v)$ when $\tau$ is small. However, this result should be taken with caution. As we can see in Figure \ref{fig:ts}.b, when $\tau$ decreases, the number of rejected classes increases and the number of $AN$-classes decreases. Indeed, the smaller the time slice, the less the behaviour between time slices is similar, which leads to a rejection of normal behaviour. $A$-classes are not affected: in most time slices, there is no node that reaches a degree within the class, whatever the time slice size. Moreover, their number increases. This is explained in Figure \ref{fig:petites_vs_gdes}. We see that for $\tau=0.25$, there is $83\%$ of $T_i$ for which the fraction $f_i(C_{34})$ is zero, against $67\%$ for $\tau=2.0$. When $\tau$ decreases, the proportion of time slices without traffic in the class compared to the ones that contain traffic is much higher than in experiments with a larger $\tau$. If the increase of $A$-classes makes it possible to identify more outliers, the decrease of $AN$-classes, on the other hand, prevents us from determining if a removal is bad or not by the appearance of a negative outlier. Hence, our validation criteria for removals cannot be applied which could lead to a disruption of normal traffic.\\ 

\begin{figure}
\hspace*{1.6cm}
\resizebox{0.33\textwidth}{!}{\input{time_slice_petites_gdes.tex}}

\vspace*{0.25cm}
\scalebox{0.85}{
\begin{tabular}{|c|c|c|}
\cline{2-3}
 \multicolumn{1}{c|}{} &  \multicolumn{2}{c|}{$\tau$}\\
\cline{2-3}
 \multicolumn{1}{c|}{} &  0.25 & 2.0 \\
  \hline
\small{Detected} & \small{$(C_{34},T_{5256}),(C_{34},T_{5257})$} & \small{$(C_{34},T_{657})$} \\
  \small{outliers} & \small{$(C_{34},T_{5258}),(C_{34},T_{5259})$} &  \multicolumn{1}{c|}{} \\
    \hline
\small{Identified} & \small{$\{(t,v) : t\in [1314,1315[ \, $} & \small{$\{(t,v) : t\in [1314,1316[ \,$}\\
 \small{outliers} & \small{$\text{ and } \, d_t(v)\in C_{34}\}$} & \small{$ \text{ and } \, d_t(v)\in C_{34}\}$}\\
    \hline
    \small{Removal} & \small{$u$ from $1314.24$ to $1314.95$} & \small{$u$ from $1314.24$ to $1314.95$} \\
        \hline
\end{tabular}
}
\vspace*{0.1cm}
\caption{\small{\textbf{Number of $A$-classes and removals depending on $\tau$ -} Node $u$ has abnormal traffic within degree class $C_{34}$. In the sample, there are $4$ time slices with abnormal traffic out of $24$ using $\tau=0.25$ and $1$ time slice with abnormal traffic out of $3$ using $\tau=2.0$. This influences the proportion of $A$-classes but not the temporal precision on which the outlier is removed: with $\tau=0.25$ (red dots), we remove $u$ from $1314.24$ to $1314.25$ ($T_{5256}$), then from $1314.25$ to $1314.75$ ($T_{5257}$,$T_{5258}$), and finally from $1314.75$ to $1314.95$ ($T_{5259}$). With $\tau=0.25$ (black triangles), we remove $u$ from $1314.24$ to $1314.95$ ($T_{657}$).}}
\label{fig:petites_vs_gdes}
\end{figure}

In addition, we see in Figure \ref{fig:ts}.c that using small time slices significantly increases computation time.\\

Finally, note that, thanks to the modelling of traffic as a link stream, the time slice size does not affect the temporal precision with which we identify events since the time period during which a node is within an $A$-class is the same whatever the considered time slice (see Figure \ref{fig:petites_vs_gdes}).

\subsection{Variations of the Degree Class Size}
\label{sec:construct_class}

We divide local degree distributions into degree classes of size $r$. The success of our method is based on the fact that distributions of fractions $f_i(C_j)$ on all time slices $T_i$ and degree classes $C_j$  are homogeneous. \\

\begin{figure*}
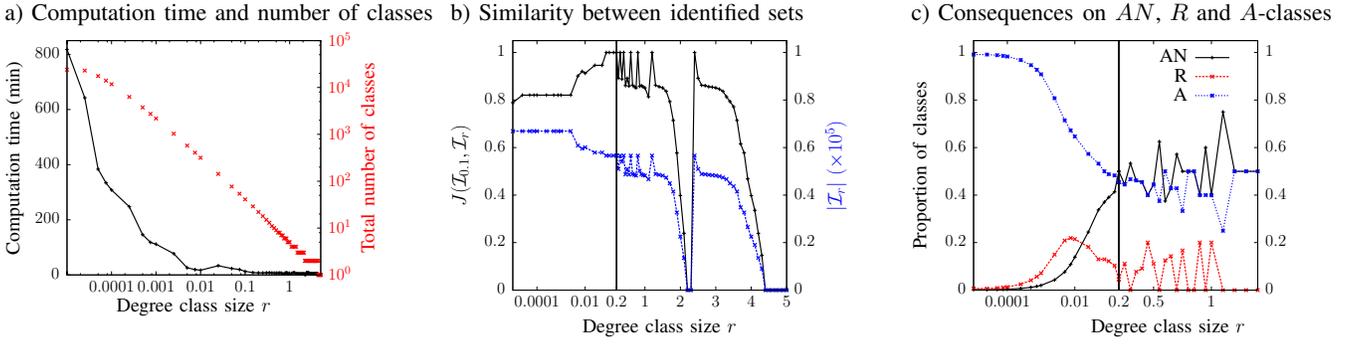

	\begin{subfigure}[t]{0.32\textwidth}
				\small{a) Computation time and number of classes}

	    \resizebox{0.95\textwidth}{!}{\input{time_classes.tex}}
	    	    	\end{subfigure}	
	\begin{subfigure}[t]{0.32\textwidth}
	\small{b) Similarity between identified sets}

	  \resizebox{0.95\textwidth}{!}{\input{jaccard_classes.tex}}
	\end{subfigure}\hspace*{0.2cm}
		\begin{subfigure}[t]{0.32\textwidth}
			\small{c) Consequences on $AN$, $R$ and $A$-classes}

	  \resizebox{0.95\textwidth}{!}{\input{proportions_AN_R_A_classes.tex}}
	\end{subfigure}  
	    	    	\vspace*{0.3cm}
\caption{\small{\textbf{Degree classes size influence -} (a) Due to the large number of classes, the computation time significantly increases as $r$ decreases. (b) The Jaccard index between identified sets shows that our method is stable from $r=10^{-5}$ to $r=1.6$. (c) Small degree classes lead to an increase of the number of $A$-classes and to a decrease of the number of $AN$-classes. For $r<0.02$, the proportion of $R$-classes is higher than the one of $AN$-classes. For $r>0.2$, proportions of $A$ and $AN$-classes are similar and the proportion of $R$-classes is low. Note that we did not plotted the proportion of classes for $r>1.4$ because of fluctuations due to the small number of classes.}}
\label{fig:taille_r}
\end{figure*}

First notice that due to class construction, the total number of classes is very large when $r$ is small and decreases rapidly when $r$ increases (see Figure \ref{fig:taille_r}.a). For $r=10^{-5}$, there are $23,983$ classes; for $r>1.6$, the total number of classes is smaller than $4$ and reaches 1 for $r\geqslant 4.4$. Consequently, the smaller $r$, the higher the computation time. \\

When we look at the size and similarity of identified sets, we observe several phenomena (see Figure \ref{fig:taille_r}.b):\\
1) the number of identified outliers increases for $r<0.2$; \\
2) we do not identify outliers for $r\in[2.2,2.3]$ and $r>4.4$;\\
3) the Jaccard index is higher than 0.8 for $r\in [10^{-5},1.6]$ and $r\in [2.4,3.3]$;\\
4) the Jaccard index fluctuates between $0.8$ and $1$ for $r\in [0.2,1.6]$;  \\
5) the number of identified outliers drops from $r=1.7$, increases from $r=2.4$, then drops again from $r=3.4$.\\
Once again, these observations are linked to the proportions of the three classes types. We explain them in detail in the following.\\

\begin{figure}[ht]
\hspace*{0.3cm}
\scalebox{0.6}{
\input{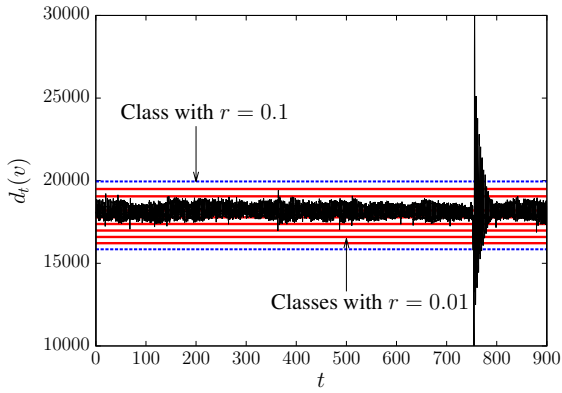}
}
\caption{\small{\textbf{Consequences of small classes -} When $r=0.01$, classes are too small and do not contain fluctuations. As a consequence, for $r=0.01$, we observe 6 $AN$-classes and 4 $R$-classes instead of only one $AN$-class for $r=0.1$. Note that this node comes from the dataset of November 2018.}}
\label{fig:gros_degre}
\end{figure}

When the degree class size is too small, classes do not longer integrates degree fluctuations (see Figure \ref{fig:gros_degre}). As we can see in Figure \ref{fig:taille_r}.c, this leads to an increase of $A$ and $R$-classes at the expense of $AN$-classes which decrease. As above, while the increase of $A$-classes makes it possible to identify more outliers (observation 1), the decrease of $AN$-classes prevents us from using our validation criteria for removals which could lead to a disruption of normal traffic.\\

\begin{figure}
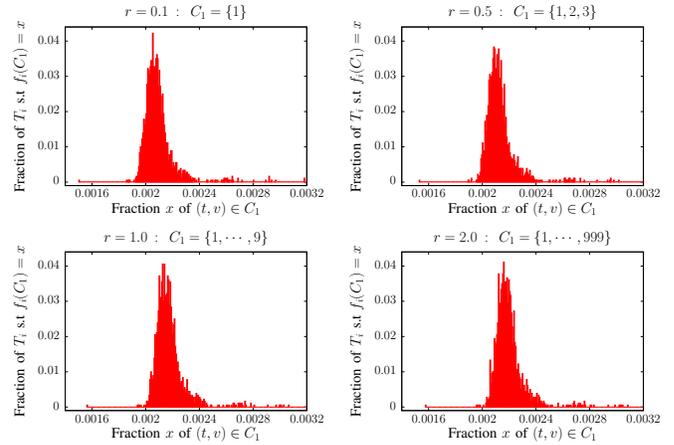

	\begin{subfigure}[t]{0.24\textwidth}
	  \resizebox{0.95\textwidth}{!}{\input{classes_1_1.tex}}
	\end{subfigure}
	\begin{subfigure}[t]{0.24\textwidth}
	    \resizebox{0.95\textwidth}{!}{\input{classes_1_3.tex}}
	    	\end{subfigure}
	
	\vspace*{0.1cm}    	
	\begin{subfigure}[t]{0.24\textwidth}
	  \resizebox{0.95\textwidth}{!}{\input{classes_1_10.tex}}
	\end{subfigure}
	\begin{subfigure}[t]{0.24\textwidth}
	    \resizebox{0.95\textwidth}{!}{\input{classes_1_100.tex}}
	    	    	\end{subfigure}	    	
	    	    	
\caption{\small{\textbf{Class $C_1$ depending on $r$ -} There are numerous couples $(t,v)$ which have degree $1$: when $C_1$ also contains degrees larger than 1, we see no difference between the corresponding distributions. Therefore, for $r>0.3$, we detect the same outliers as with $r=0.1$, since outliers for which $d_t(v)>1$ are included within the Gaussian. Note that we zoomed on the Gaussian for greater clarity.}}
	    	\label{fig:no_diff}
\end{figure}

On the contrary, when the degree class size is too large, classes integrate too much traffic. As a consequence, we can see in Figure \ref{fig:taille_r}.c, that the more $r$ increases the more $A$-classes decreases. Therefore, the more $r$ increases, the less we identify outliers. This is how we explain observation 2: for $r=2.2$, there are two $AN$-classes; for $r=2.3$; there are one $AN$-class and one $R$-class; and finally, for $r\geqslant 4.4$, there is only one $AN$-class. Moreover, we see in Figure \ref{fig:no_diff} that when class $C_1$ contains several values of degrees, the resulting distribution looks the same as when it only contains degree $1$. Hence, the large proportion of couples $(t,v)$ having degree $1$ obstructs the traffic of couples $(t,v)$ which have a degree larger than $1$. As a consequence, we detect less outliers and incorrect removals could be accepted which could also lead to a disruption of normal traffic. \\

\begin{figure}[ht]
\hspace*{0.3cm}
\scalebox{0.6}{
\input{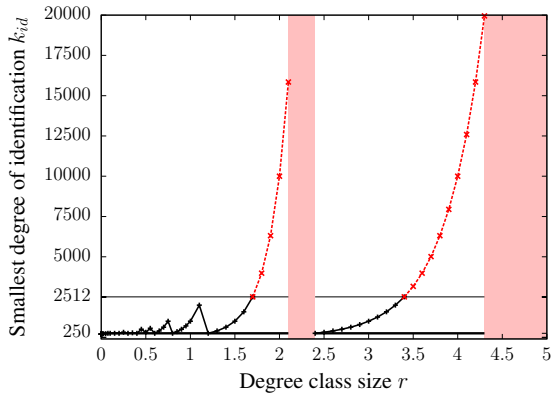}
}
\caption{\small{\textbf{Smallest degree of identification according to $r$ -} When $k_{id}\in[250,2512]$, the Jaccard index is higher than 0.8. However, when $k_{id} > 2512$, the number of identified outliers drops (dashed red line) and reaches $0$ in shaded zones. }}
\label{fig:dispo}
\end{figure}

Finally, observations 3, 4 and 5 are explained by discretization effects. In the one-hour traffic trace, classes are arranged such that low degree classes are $AN$-classes and high degree classes are $A$-classes. Let $k_{id}$ be the smallest degree from which we are able to identify events. As we can see in Figure \ref{fig:dispo}, $k_{id}$ is different depending on the chosen $r$. The smaller $k_{id}$, the larger the number of detected outliers. Until $r=0.2$, $k_ {id}$ is lower than $250$ and the number of identified outliers is maximal. Then, $k_{id}$ fluctuates between $250$ and $2,512$, which explain observation 4. A lot of outliers are located within this range. Hence, if $k_{id}\in [250,2512]$, these outliers are identified, otherwise they are not, which explains observation 3. Regarding observation 5, this is what happens: for $ r \in [1.7,2.1] $, the number of classes is equal to three. There are two $AN$-classes and one $A$-class. The smallest degree of identification $k_{id}$ increases with $r$ which causes the drop in the number of identified outliers. For $r\in[2.2,4.3]$, the total number of classes is two. The number of detected outliers depends on classes proportions: there are either two $AN$-classes ($ r = 2.2 $), or one $AN$-class and one $R$-class ($r=2.3$), or one $AN$-class and one $A$-class ($ r \in [2.4,4.3] $). Finally, we observe the same phenomenon, for $r\in [2.4,4.3]$, $k_{id} $ increases with $r$ which causes the drop of the number of identified outliers until only one class remains for $ r>4.3$. \\

The method instability with respect to r is only observed for a number of classes lower than 4 ($ r> 1.6 $). For a number of classes ranging from $24,000$ ($r=10^{-5}$) to $4$, the method is stable and exhibits very similar results. Therefore, $r$ must be chosen based on data range, by keeping a reasonably high number of classes.

\section{Conclusion}
\label{sec:conclusion}
When dealing with IP traffic, we are faced with IP addresses having very different behaviours. In this context, one question arises: how to differentiate normal behaviours from abnormal ones. In this paper, we proposed a solution to this issue. We introduced a method that detects outliers in IP traffic modelled as a link stream by studying the degree of node over time. We applied our method on three datasets from the MAWI archive: one-hour long IP traffic trace from June 2013, one-day long IP traffic trace from June 2013 and a fifteen-minute long IP traffic trace from November 2018. Likewise, we performed series of experiments by varying the parameters used. In all these situations, we obtained stable results pointing interesting anomalous activities in IP traffic, independently of the degree order of magnitude. Moreover, we surgically removed anomalous traffic, which allowed us to validate our identification, identify more subtle outliers and recover a normal traffic with respect to the degree feature.\\
This work however is only a first step towards anomaly detection in link streams and may be improved on several aspects. We could extend our method with more complex features than the degree in order to find more complex anomalies and to identify the remaining events unidentified with the degree. This task would be simplified by the fact that largest anomalies have already been removed from the remaining traffic, allowing for a more detailed and finer analysis. At broader scale, our work could be useful in the field of IP traffic modelling as we would be able to generate normal traffic according to a specific feature. Likewise, thanks to their individual extraction, anomalies could also be studied separately for a better characterization.

\section*{Acknowledgement}
This work is funded in part by the European Commission H2020 FETPROACT 2016-2017 program under grant 732942 (ODYCCEUS), by the ANR (French National Agency of Research) under grants ANR-15- E38-0001 (AlgoDiv), by the Ile-de-France Region and its program FUI21 under grant 16010629 (iTRAC).\\

\bibliographystyle{abbrv}
\bibliography{biblio_article_ip.bib}


\end{document}